\documentclass[11pt]{article}
\usepackage{moriond,epsfig}

\def\tt{$t\bar{t}~$}
\def\Journal#1#2#3#4{{#1} {\bf #2}, #3 (#4)}

\def\PRL{\em Phys. Rev. Lett.}
\def\PRD{{\em Phys. Rev.} D}

\def\be{\begin{equation}}
\def\ee{\end{equation}}
\def\bea{\begin{eqnarray}}
\def\eea{\end{eqnarray}}

\begin{document}
\vspace*{4cm}
\title{Top quark physics at LHC}

\author{ M. Vander Donckt,\\
 on behalf of CMS and ATLAS Collaborations}

\address{CERN, 1211 Gen\`eve 23, Switzerland}

\maketitle\abstracts{
The Large Hadron Collider (LHC) will provide a huge amount of top-antitop events, making the LHC a top quark factory, producing 1 \tt pair per second at a luminosity of $10^{33}cm^{-2}s^{-1}$. A large top quark sample will be available from the start of LHC and will play an important role in commissioning the CMS and ATLAS detectors. An overview of the top quark measurements during the first data-taking period is given.}

\section{Introduction}
The Large Hadron Collider (LHC) will start to collide protons in July 2008. The large amount of \tt pairs produced will
provide the opportunity to measure many top properties with a precision never reached before. The LHC
can be considered as a real top quark factory as eight (three) millions of \tt pairs (single tops)
will be produced per year during its first phase, corresponding to an initial luminosity of $10^{33} cm^{-2}s^{-1}$.
Top quark pair production at the LHC happens mainly via gluon fusion. Final states result from the decay of two top quarks into Wb, with a branching fraction close to unity, and the subsequent decay of the W boson. The
resulting channels are therefore called fully hadronic (46\%), semi-leptonic (44\%) and fully leptonic (10\%), depending on the W decays.\\
 Both ATLAS \cite{atlasTDR} and CMS \cite{cmsTDR} have conducted extensively studies in the
top quark sector, trying to search for new physics by probing the Standard Model. Moreover, \tt events provide an excellent
environment for calibration of the jet energy scale and b-tagging algorithms.

\section{Early top signal}
\begin{figure}[hhh]
\begin{center}
\epsfig{file=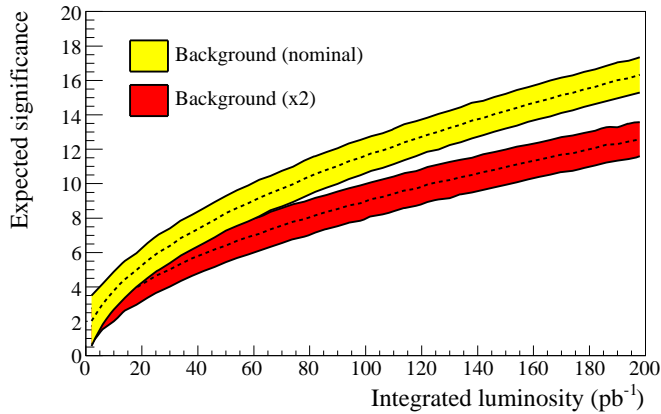,width=0.6\textwidth,clip}
\caption{ATLAS Top signal significance as a function of the accumulated statistics, for W+jets background and twice this background to account for QCD background.}
\label{fig.topsig}
\end{center}
\end{figure}

The top quark pair production cross section will be one of the
first relevant measurements to be performed. The main challenge is to select a pure top quark sample without the application of b-tagging identification algorithms.ATLAS expects to isolate \tt events without requesting b-tagging within the first $100pb^{-1}$ of data. The selection of \tt pairs in the semi-leptonic events is performed by requesting a lepton with $p_t>20GeV/c$, 3 jets with $E_T>40GeV$, 1 jet with  $E_T>20GeV$ and missing $E_T>20GeV/c$.  
The expected signal significance as a function of the integrated luminosity is shown in figure \ref{fig.topsig}.

\section{Commissioning with top quarks}
Top quark events at the LHC are very complex and their reconstruction is based on almost each crucial aspect of the
detector. Top quark are therefore perfect for the commissioning of the detectors and measuring the physics performance of the experiments at start-up. \\
The measurement of the jet energy scale calibration factors \cite{2006-25} can be extracted from the events collected in the first
weeks of data taking: the W boson mass constraint can be used to reach an uncertainty of better than 1\% with an integrated luminosity of about $300 pb^{-1}$.
The main challenge is the control of the pile-up collisions, for which vertex counting methods can be envisaged.\\
A tag and probe method is used by CMS\cite{2006-13} to measure the b-tagging efficiency with semi-leptonic \tt pairs. The b-tagged jet associated to the hadronic W is used as a tag while the jet on the leptonic side is probed. A 6\% (10\%)uncertainty can be reached in the barrel (endcaps) for an integrated luminosity of 1$fb^{-1}$.
\begin{figure}[hhh]
\begin{center}
\epsfig{file=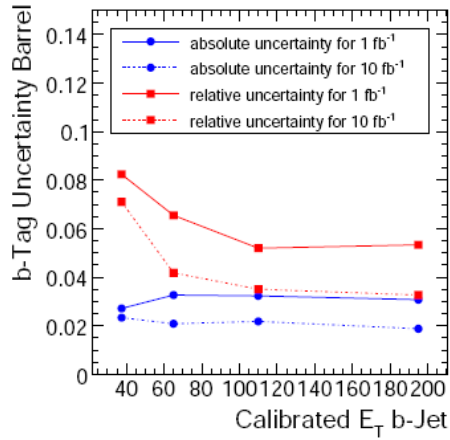,width=0.4\textwidth,clip}
\end{center}
\caption{CMS barrel uncertainty of the b-tagging efficiency as a function of the b-jet $E_T$.}
\label{fig.btag}
\end{figure}

\section{Cross-section and mass measurements}
Several methods have been explored for different final state topologies and the studies have shown that the expected uncertainty, for an integrated luminosity of 10 $fb^{-1}$, on the mass measurement could reach this goal if detectors and theoretical systematic uncertainty are well under control \cite{atlas-topmass,cms-ttsemilmass,cms-ttother}. 
A similar program for the cross section also exists, but
the systematic uncertainty is at present limited to  10\% because of the currently relatively large uncertainties on
b-tagging algorithms and the luminosity.\\
Among the large list of systematic contributions that enter the mass and cross-section measurements, the major
effects (depending on the decay channel) are the light and b-jet energy scale, the initial/final state radiation modeling, the minimum-bias and underlying events estimate, the parton distribution functions, the uncertainty on b-tag performances.\\
A 1.5\% uncertainy is needed on the b-jet energy scale calibration to reach the precision of 1 $GeV/c^2$ on the top quark mass using the semi-leptonic \tt events. Most of the systematic effects have to be extracted from independent data samples to have a robust and reliable evaluation. Both ATLAS and CMS have a detailed program during the initial period of data taking, to measure these effects directly from data.
\section{Spin correlations}
Unlike light quarks, the top quark decays before hadronization or depolarization can take place. The study of the angular correlation between the decay products (b-jet, lepton or light quark from W) in the top (anti-top) restframe and the top (anti-top) direction in the \tt restframe gives the unique opportunity to study the top quark spin.  A CMS analysis\cite{cms-spin} computes that the correlation coefficient can be measured with a total relative uncertainty of 17\% with $10fb^{-1}$ of semi-leptonic \tt data. The main sources of systematic uncertainties for this analysis are the b-tagging efficiency and the jet energy scale and multiplicity. ATLAS \cite{atlas-spin} combines di-leptonic and semi-leptonic analyses for $10fb^{-1}$ to reach a 4\% precision for Standard Model spin correlations. 
\section{Single top production}
Production of top quark events at LHC is not limited to \tt pairs. Indeed, single top production via the s and t
channel, as well as the associated production of a top quark and a W boson, account for a third of the overall
production cross-section of top quarks at the LHC. The NLO total cross section for each process amounts to
10 pb, 250 pb and 70 pb respectively \cite{xsec-singletop}. Because of the different final states and topologies, different
selection strategies have to be developed and different analysis will be performed both in ATLAS\cite{atlas-singletop} and CMS\cite{cms-singletop}.
The general systematic considerations accounted for the \tt~mass and cross-section measurements apply also
to measurements with single top quark events.
A detailed overview of the CMS analysis for cross section measurement in single top quark production, reported
in \cite{cms-singletop}, shows that the t-channel analysis appears as the most favourable case due to its higher production cross
section and better signal over background ratio. The precision that can be achieved with an integrated luminosity
of 10 $fb^{−1}$ is: $\frac{\Delta\sigma}{\sigma}=2.7\% (stat) \pm 8.1\% (syst) \pm 3\%(luminosity)$
The measurement of the cross sections are directly related to the $|V_{tb}|$ matrix element in the CKM matrix. If
the SM holds, stringent limits on couplings of new interactions can be established using the measurement of the
t-channel cross section. 
\section{Search for new physics}
\begin{figure}[hhh]
\begin{center}
\epsfig{file=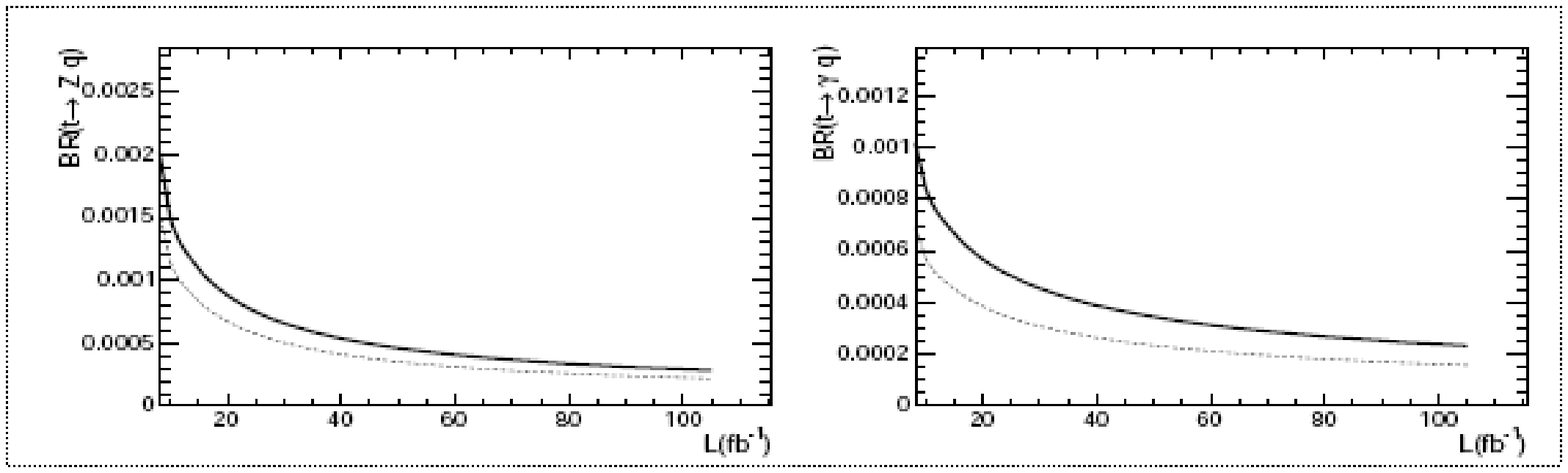,width=0.9\textwidth,clip}
\end{center}
\caption{CMS branching ratios of FCNC top decays including (solid line) and excluding (dashed line) the contribution from systematic uncertainties as a function of integrated luminosity, for $t\rightarrow qZ$ (left) and $t\rightarrow q\gamma$ (right), assuming a 5-sigma discovery level for an integrated luminosity of L=10$fb^{-1}$.}
\label{fig.cmsfcnc}
\end{figure}

Flavour changing neutral currents (FCNC) are strongly suppressed in the Standard Model.In the top quark sector of the
SM, these contributions limit the FCNC decay branching ratios to the gauge bosons,
BR($t\rightarrow qX$) (X = Z;$\gamma$,g), to below $10^{-10}$. Both ATLAS\cite{atlas-fcnc} and CMS\cite{cms-fcnc} have analyses ready to search for FCNC in top quark decays as possible hints for physics beyond the Standard Model. The expected sensitivity reach with 10$fb^{-1}$ (shown in figure \ref{fig.cmsfcnc}) extends by 2 orders of magnitudes that of the Tevatron \cite{cdf-fcnc}.

\section{Conclusions}
The LHC will open a new era of precision measurements in the top quark physics that will lead to a thorough determination of the top quark properties.  CMS and ATLAS are ready to exploit the first top events to commission the detectors. This will be a first step towards stringent tests of the SM which are powerful probes in the search for new physics.

\end{document}